\documentclass[12pt]{article}
\begin{document}

\begin{center}
{\Large\bf{}Algebraic Rainich conditions for the tensor V}
\end{center}

\begin{center}
Lau Loi So\\
Department of Physics, National Central University, Chung Li 320, Taiwan\\
Department of Physics, Tamkang University, Tamsui 251, Taiwan\\
(s0242010@gmail.com, dated on 5 May 2010)
\end{center}

\begin{abstract}
Algebraic conditions on the Ricci tensor in the
Rainich-Misner-Wheeler unified field theory are known as the
Rainich conditions. Penrose and more recently Bergqvist and
Lankinen made an analogy from the Ricci tensor to the Bel-Robinson
tensor $B_{\alpha\beta\mu\nu}$, a certain fourth rank tensor
quadratic in the Weyl curvature, which also satisfies algebraic
Rainich-like conditions.  However, we found that not only does the
tensor $B_{\alpha\beta\mu\nu}$ fulfill these conditions, but so
also does our recently proposed tensor $V_{\alpha\beta\mu\nu}$,
which has many of the desirable properties of
$B_{\alpha\beta\mu\nu}$. For the quasilocal small sphere limit
restriction, we found that there are only two fourth rank tensors
$B_{\alpha\beta\mu\nu}$ and $V_{\alpha\beta\mu\nu}$ which form a
basis for good energy expressions. Both of them have the
completely trace free and causal properties, these two form
necessary and sufficient conditions. Surprisingly either
completely traceless or causal is enough to fulfill the algebraic
Rainich conditions. Furthermore, relaxing the quasilocal
restriction and considering the general fourth rank tensor, we
found two remarkable results: (i) without any symmetry
requirement, the algebraic Rainich conditions only require totally
trace free; (ii) with a symmetry requirement, we recovered the
same result as in the quasilocal small sphere limit.
\end{abstract}

\section{Introduction}
In 1925 Rainich proposed a unified field theory for source-free
electromagnetism and gravitation \cite{Rainich}.  Misner and
Wheeler \cite{MisnerWheeler} 32 years later proposed a
geometrically unified theory, based on the Rainich idea, now
called the Rainich-Misner-Wheeler theory. The necessary conditions
are called the Rainich conditions.  The algebraic Rainich
conditions refer to the Ricci tensor, but this tensor can be
replaced by other tensors.  Penrose \cite{Penrose} and more
recently Bergqvist and Lankinen
\cite{BergqvistLankinen,BergqvistLankinen2nd} made an analogy from
the Ricci tensor to the Bel-Robinson tensor
$B_{\alpha\beta\mu\nu}$, a certain fourth rank tensor quadratic in
the Weyl curvature, which also satisfies algebraic Rainich-like
conditions.  We found that not only does the tensor
$B_{\alpha\beta\mu\nu}$ fulfill these conditions, but so also does
our recently proposed tensor $V_{\alpha\beta\mu\nu}$, which has
many of the desirable properties of $B_{\alpha\beta\mu\nu}$.

The Bel-Robinson tensor possesses many nice properties; it is
completely symmetric, totally trace-free and divergence-free. It
also satisfies the dominant energy condition \cite{Senovilla},
\begin{equation}
B_{\alpha\beta\mu\nu}u^{\alpha}v^{\beta}w^{\mu}z^{\nu}\geq{}0,
\end{equation}
for any timelike unit normal vectors $u$, $v$, $w$ and $z$. An
unique alternative tensor $V_{\alpha\beta\mu\nu}$ was proposed
recently which gives the same non-negative gravitational energy
density in the small sphere limit \cite{SoarXiv2009}
\begin{equation}
V_{\alpha\beta\mu\nu}:=S_{\alpha\beta\mu\nu}+K_{\alpha\beta\mu\nu}
\equiv{}B_{\alpha\beta\mu\nu}+W_{\alpha\beta\mu\nu},\label{16zJan2010}
\end{equation}
where $B_{\alpha\beta\mu\nu}$, $S_{\alpha\beta\mu\nu}$,
$K_{\alpha\beta\mu\nu}$ and $W_{\alpha\beta\mu\nu}$ are defined in
section 2. While $V_{\alpha\beta\mu\nu}$
\cite{SoarXiv2009,SoCQG2009} does not have the completely
symmetric property, it does fulfill the totally trace free
property and satisfies the weak energy condition,
\begin{equation}
V_{\alpha\beta\mu\nu}u^{\alpha}u^{\beta}u^{\mu}u^{\nu}\equiv
{}B_{\alpha\beta\mu\nu}u^{\alpha}u^{\beta}u^{\mu}u^{\nu}\geq{}0.
\end{equation}

The algebraic Rainich conditions \cite{Rainich,Stephani} are
\begin{eqnarray}
R_{\alpha\lambda}R_{\beta}{}^{\lambda}=\frac{1}{4}g_{\alpha\beta}R_{\rho\lambda}R^{\rho\lambda},\quad\quad{}
R^{\lambda}{}_{\lambda}=0,\quad\quad{}
R_{\alpha\beta}u^{\alpha}u^{\beta}\geq{}0,
\end{eqnarray}
where $R_{\alpha\beta}$ is the second rank Ricci tensor.  Turning
to higher rank, one can use the Bel-Robinson tensor which is the
first fourth rank tensor people recognized that satisfies these
algebraic Rainich conditions \cite{Penrose,BergqvistLankinen}. It
is known that the Ricci tensor is symmetric; if we make an analogy
from second rank to fourth rank, the completely symmetric property
need not be required.  Therefore, as far as the quasilocal small
sphere limit is concerned, we found the unique alternative fourth
rank tensor $V_{\alpha\beta\mu\nu}$ which satisfies the algebraic
Rainich conditions.

Interestingly, we discovered that
$X_{\alpha\lambda\sigma\tau}Y_{\beta}{}^{\lambda\sigma\tau}
=\frac{1}{4}g_{\alpha\beta}X_{\rho\lambda\sigma\tau}Y^{\rho\lambda\sigma\tau}$,
where $X_{\alpha\beta\mu\nu}$ and $Y_{\alpha\beta\mu\nu}$ are any
quadratic in Riemann curvature tensors.  This indicates that this
is an identity (i.e., not restricted to the quasilocal small
sphere limit) which means it is no longer a condition. Therefore
the algebraic Rainich conditions for fourth rank leave two
conditions, not the expected three.

Under the quasilocal small sphere limit restriction, we found that
there are only two fourth rank tensors $B_{\alpha\beta\mu\nu}$ and
$V_{\alpha\beta\mu\nu}$ forming a basis for good expressions. Both
of them have the completely trace free and causal properties, we
found that these two properties form necessary and sufficient
conditions.  In other words, either completely traceless or causal
can fulfill the algebraic Rainich conditions.

Furthermore, relaxing the quasilocal restriction and considering
the general fourth rank tensor, we found two important results.
One is without any symmetry requirement: we found that the
algebraic condition only requires completely traceless.  The other
is imposing some certain symmetry: we recovered, as expected, the
same result as in the quasilocal small sphere limit.

\section{Technical background}
The Bel-Robinson tensor \cite{BelRobinson} was proposed in 1958 as
a certain quadratic combination of the Weyl tensor:
\begin{eqnarray}
B_{\alpha\beta\mu\nu}&:=&R_{\alpha\lambda\mu\sigma}R_{\beta}{}^{\lambda}{}_{\nu}{}^{\sigma}
+\ast{}R_{\alpha\lambda\mu\sigma}\ast{}R_{\beta}{}^{\lambda}{}_{\nu}{}^{\sigma}\nonumber\\
&=&R_{\alpha\lambda\mu\sigma}R_{\beta}{}^{\lambda}{}_{\nu}{}^{\sigma}
+R_{\alpha\lambda\nu\sigma}R_{\beta}{}^{\lambda}{}_{\mu}{}^{\sigma}
-\frac{1}{2}g_{\alpha\beta}R_{\mu\lambda\sigma\tau}R_{\nu}{}^{\lambda\sigma\tau},\label{15yJan2010}
\end{eqnarray}
where $\ast{}R_{\alpha\lambda\sigma\tau}$ is the dual of
$R_{\alpha\lambda\sigma\tau}$. One place where the Bel-Robinson
tensor naturally shows up is in the expressions for gravitational
energy in a small region. There are three fundamental tensors that
commonly occur in the gravitational pseudotensor expressions
\cite{SoIJMPD,PRD} in vacuum:
\begin{eqnarray}
B_{\alpha\beta\mu\nu}
&:=&R_{\alpha\lambda\mu\sigma}R_{\beta}{}^{\lambda}{}_{\nu}{}^{\sigma}
+R_{\alpha\lambda\nu\sigma}R_{\beta}{}^{\lambda}{}_{\mu}{}^{\sigma}
-\frac{1}{8}g_{\alpha\beta}g_{\mu\nu}\mathbf{R}^{2},\label{16aJan2010}\\
S_{\alpha\beta\mu\nu}
&:=&R_{\alpha\mu\lambda\sigma}R_{\beta\nu}{}^{\lambda\sigma}
+R_{\alpha\nu\lambda\sigma}R_{\beta\mu}{}^{\lambda\sigma}
+\frac{1}{4}g_{\alpha\beta}g_{\mu\nu}\mathbf{R}^{2},\label{29April2009}\\
K_{\alpha\beta\mu\nu}
&:=&R_{\alpha\lambda\beta\sigma}R_{\mu}{}^{\lambda}{}_{\nu}{}^{\sigma}
+R_{\alpha\lambda\beta\sigma}R_{\nu}{}^{\lambda}{}_{\mu}{}^{\sigma}
-\frac{3}{8}g_{\alpha\beta}g_{\mu\nu}\mathbf{R}^{2}.\label{16bJan2010}
\end{eqnarray}
where $\mathbf{R}^{2}=R_{\rho\tau\xi\kappa}R^{\rho\tau\xi\kappa}$
and we have rewritten (\ref{15yJan2010}) by substituting the well
known identity \cite{Yefremov} in vacuum
\begin{equation}
R_{\mu\lambda\sigma\tau}R_{\nu}{}^{\lambda\sigma\tau}
\equiv\frac{1}{4}g_{\mu\nu}\mathbf{R}^{2}.\label{15wJan2010}
\end{equation}
On the other hand, we also introduced another tensor
$W_{\alpha\beta\mu\nu}$ which gives an alternative representation
of $V_{\alpha\beta\mu\nu}$ as denoted in (\ref{16zJan2010})
\begin{equation}
W_{\alpha\beta\mu\nu}:=\frac{3}{2}S_{\alpha\beta\mu\nu}-\frac{5}{8}g_{\alpha\beta}g_{\mu\nu}\mathbf{R}^{2}
+\frac{1}{8}(g_{\alpha\mu}g_{\beta\nu}+g_{\alpha\nu}g_{\beta\mu})\mathbf{R}^{2},
\end{equation}

The analog of the electric part $E_{ab}$ and magnetic part
$H_{ab}$ are defined in terms of the Weyl tensor \cite{Carmeli} as
follows
\begin{equation}
E_{ab}:=C_{a0b0},\quad{}H_{ab}:=*C_{a0b0},\quad{} a,b=1,2,3.
\end{equation}
The fundamental property of a tensor is that if it vanishes in one
frame, then it must vanish in any other frame. This is an
elementary property for a tensor, however, remarkably it provides
an easy and efficient way for the verification of identities. For
instance, we have checked using orthonormal frames that the
identity in (\ref{15wJan2010}) is true.

Here are some properties of $S_{\alpha\beta\mu\nu}$,
$K_{\alpha\beta\mu\nu}$, $W_{\alpha\beta\mu\nu}$ and
$V_{\alpha\beta\mu\nu}$ that we found:
\begin{eqnarray}
S_{\alpha\beta\mu\nu}&\equiv&S_{(\alpha\beta)(\mu\nu)}\equiv{}S_{(\mu\nu)(\alpha\beta)},\\
S_{\alpha\beta\mu}{}^{\mu}&\equiv&\frac{1}{4}g_{\alpha\beta}S_{\rho}{}^{\rho}{}_{\mu}{}^{\mu}
\equiv{}\frac{3}{2}g_{\alpha\beta}\mathbf{R}^{2},\quad{}
S_{\alpha\mu\beta}{}^{\mu}\equiv\frac{1}{4}g_{\alpha\beta}S_{\rho\mu}{}^{\rho\mu}
\equiv{}0,\label{9aSep2009}\\
K_{\alpha\beta\mu\nu}&\equiv&K_{(\alpha\beta)(\mu\nu)}\equiv{}K_{(\mu\nu)(\alpha\beta)},\\
K_{\alpha\beta\mu}{}^{\mu}&\equiv&\frac{1}{4}g_{\alpha\beta}K_{\rho}{}^{\rho}{}_{\mu}{}^{\mu}
\equiv{}-\frac{3}{2}g_{\alpha\beta}\mathbf{R}^{2},\quad
{}K_{\alpha\mu\beta}{}^{\mu}\equiv\frac{1}{4}g_{\alpha\beta}K_{\rho\mu}{}^{\rho\mu}\equiv{}0,
\label{9bSep2009}\\
W_{\alpha\beta\mu\nu}&\equiv&W_{(\alpha\beta)(\mu\nu)}\equiv{}W_{(\mu\nu)(\alpha\beta)},\\
W_{\alpha\beta\mu}{}^{\mu}&\equiv&\frac{1}{4}g_{\alpha\beta}W_{\rho}{}^{\rho}{}_{\mu}{}^{\mu}
\equiv{}0,\quad
{}W_{\alpha\mu\beta}{}^{\mu}\equiv\frac{1}{4}g_{\alpha\beta}W_{\rho\mu}{}^{\rho\mu}\equiv{}0,\\
V_{\alpha\beta\mu\nu}&\equiv&V_{(\alpha\beta)(\mu\nu)}\equiv{}V_{(\mu\nu)(\alpha\beta)},\\
V_{\alpha\beta\mu}{}^{\mu}&\equiv&\frac{1}{4}g_{\alpha\beta}V_{\rho}{}^{\rho}{}_{\mu}{}^{\mu}
\equiv{}0,\quad{}
V_{\alpha\mu\beta}{}^{\mu}\equiv\frac{1}{4}g_{\alpha\beta}V_{\rho\mu}{}^{\rho\mu}
\equiv{}0.
\end{eqnarray}
Note that, unlike the Bel-Robinson tensor, both
$S_{\alpha\beta\mu\nu}$ and $K_{\alpha\beta\mu\nu}$ are neither
totally symmetric nor totally trace free \cite{SoCQG2009}.

\section{Quadratic tensor identities for $B$, $S$,
$K$, $V$ }

Owing to the equivalence principle, gravitational energy cannot be
detected at a point.  Therefore we use quasilocal methods
(including pseudotensors).  Dealing with the quasilocal small
sphere limit approximation, consider all the possible combinations
of the small region energy-momentum density in vacuum; the general
expression is \cite{SoCQG}
\begin{equation}
2\kappa\,t_{\alpha}{}^{\beta}=2G_{\alpha}{}^{\beta
}+(a_{1}\tilde{B}_{\alpha}{}^{\beta}{}_{\mu\nu}
+a_{2}\tilde{S}_{\alpha}{}^{\beta}{}_{\mu\nu}
+a_{3}\tilde{K}_{\alpha}{}^{\beta}{}_{\mu\nu}
+a_{4}\tilde{T}_{\alpha}{}^{\beta}{}_{\mu\nu})x^{\mu}x^{\nu}
+{\cal{}O}(\mbox{Ricci},x)+{\cal{}O}(x^{3}),\label{31aMar2010}
\end{equation}
where $\kappa=8\pi{}G/c^{4}$ (here we take units such that $c=1$
for simplicity) and $a_{1}$ to $a_{4}$ are real numbers.  For the
quadratic curvature tensors in (\ref{31aMar2010}), there are four
independent basis \cite{SoarXiv2009,Deser} expressions with
certain symmetries which we used:
\begin{eqnarray}
\tilde{B}_{\alpha\beta\mu\nu}
&:=&R_{\alpha\lambda\mu\sigma}R_{\beta}{}^{\lambda}{}_{\nu}{}^{\sigma}
+R_{\alpha\lambda\nu\sigma}R_{\beta}{}^{\lambda}{}_{\mu}{}^{\sigma}
=B_{\alpha\beta\mu\nu}+\frac{1}{8}g_{\alpha\beta}g_{\mu\nu}\mathbf{R}^{2},\label{20aJan2009}\\
\tilde{S}_{\alpha\beta\mu\nu}
&:=&R_{\alpha\mu\lambda\sigma}R_{\beta\nu}{}^{\lambda\sigma}
+R_{\alpha\nu\lambda\sigma}R_{\beta\mu}{}^{\lambda\sigma}
=S_{\alpha\beta\mu\nu}-\frac{1}{4}g_{\alpha\beta}g_{\mu\nu}\mathbf{R}^{2},\label{29aApril2009}\\
\tilde{K}_{\alpha\beta\mu\nu}
&:=&R_{\alpha\lambda\beta\sigma}R_{\mu}{}^{\lambda}{}_{\nu}{}^{\sigma}
+R_{\alpha\lambda\beta\sigma}R_{\nu}{}^{\lambda}{}_{\mu}{}^{\sigma}
=K_{\alpha\beta\mu\nu}+\frac{3}{8}g_{\alpha\beta}g_{\mu\nu}\mathbf{R}^{2},\\
\tilde{T}_{\alpha\beta\mu\nu}
&:=&-\frac{1}{8}g_{\alpha\beta}g_{\mu\nu}\mathbf{R}^{2}.\label{20bJan2009}
\end{eqnarray}
Note that none of these four tensors has the completely symmetric
property, e.g., $\tilde{B}_{0011}\neq\tilde{B}_{0101}$ in general.
However, all of them do have certain symmetries, precisely
$\tilde{X}_{\alpha\beta\mu\nu}=\tilde{X}_{(\alpha\beta)(\mu\nu)}
=\tilde{X}_{(\mu\nu)(\alpha\beta)}$.  Although there exists some
other tensors different from $\tilde{B}_{\alpha\beta\mu\nu}$,
$\tilde{S}_{\alpha\beta\mu\nu}$, $\tilde{K}_{\alpha\beta\mu\nu}$
and $\tilde{T}_{\alpha\beta\mu\nu}$, they are just linear
combinations of these four.  For instance \cite{Deser}
\begin{equation}
\tilde{T}_{\alpha\mu\beta\nu}+\tilde{T}_{\alpha\nu\beta\mu}
\equiv\widetilde{B}_{\alpha\beta\mu\nu}
+\frac{1}{2}\tilde{S}_{\alpha\beta\mu\nu}
-\tilde{K}_{\alpha\beta\mu\nu}
+2\tilde{T}_{\alpha\beta\mu\nu}.\label{7Nov2008}
\end{equation}
The above identity can be obtained by making use of the completely
symmetric property of the Bel-Robinson tensor. Using
(\ref{7Nov2008}), we can rewrite the Bel-Robinson tensor in a
different representation \cite{Deser}:
\begin{equation}
B_{\alpha\beta\mu\nu}\equiv-\frac{1}{2}S_{\alpha\beta\mu\nu}
+K_{\alpha\beta\mu\nu}+\frac{5}{8}g_{\alpha\beta}g_{\mu\nu}\mathbf{R}^{2}
-\frac{1}{8}(g_{\alpha\mu}g_{\beta\nu}+g_{\alpha\nu}g_{\beta\mu})\mathbf{R}^{2}.
\label{29Oct2008}
\end{equation}

There is a known formula for the quadratic Bel-Robinson tensor
\cite{Penrose}
\begin{equation}
B_{\alpha\lambda\sigma\tau}B_{\beta}{}^{\lambda\sigma\tau}
\equiv\frac{1}{4}g_{\alpha\beta}B_{\rho\lambda\sigma\tau}B^{\rho\lambda\sigma\tau},\label{11aJan2010}
\end{equation}
which was given by Penrose using spinor methods \cite{Penrose}. We
have verified this identity using orthonormal frames (for details
see (\ref{31bMar2010}) below), moreover, using the same method, we
found the following identity
\begin{equation}
S_{\alpha\lambda\sigma\tau}S_{\beta}{}^{\lambda\sigma\tau}
\equiv\frac{1}{4}g_{\alpha\beta}S_{\rho\lambda\sigma\tau}S^{\rho\lambda\sigma\tau}.\label{18cJan2010}
\end{equation}
This is a milestone for verifying other quadratic identities
(e.g.,
$K_{\alpha\lambda\sigma\tau}K_{\beta}{}^{\lambda\sigma\tau}$) in
an easier way. In other words, one can use the same method in
orthonormal frames to verify all the possible combinations, but it
would take much unnecessary work. Instead we used simple algebra
substitution. Remember that $B_{\alpha\beta\mu\nu}$ is completely
symmetric and trace-free; making use of (\ref{29April2009}) and
(\ref{29Oct2008}), we found
\begin{eqnarray}
0&=&S_{\alpha\lambda\sigma\tau}B_{\beta}{}^{\lambda\sigma\tau}
\equiv-\frac{1}{2}S_{\alpha\lambda\sigma\tau}S_{\beta}{}^{\lambda\sigma\tau}
+S_{\alpha\lambda\sigma\tau}K_{\beta}{}^{\lambda\sigma\tau}
+\frac{15}{16}g_{\alpha\beta}\mathbf{R}^{2}\mathbf{R}^{2},\label{18aJan2010}\\
0&=&S_{\rho\lambda\sigma\tau}B^{\rho\lambda\sigma\tau}
\equiv-\frac{1}{2}S_{\rho\lambda\sigma\tau}S^{\rho\lambda\sigma\tau}
+S_{\rho\lambda\sigma\tau}K^{\rho\lambda\sigma\tau}
+\frac{15}{4}\mathbf{R}^{2}\mathbf{R}^{2}.\label{18bJan2010}
\end{eqnarray}
Rewrite (\ref{18aJan2010}) and (\ref{18bJan2010}) as follows:
\begin{eqnarray}
S_{\alpha\lambda\sigma\tau}K_{\beta}{}^{\lambda\sigma\tau}
&\equiv&\frac{1}{2}S_{\alpha\lambda\sigma\tau}S_{\beta}{}^{\lambda\sigma\tau}
-\frac{15}{16}g_{\alpha\beta}\mathbf{R}^{2}\mathbf{R}^{2},\label{18dJan2010}\\
S_{\rho\lambda\sigma\tau}K^{\rho\lambda\sigma\tau}
&\equiv&\frac{1}{2}S_{\rho\lambda\sigma\tau}S^{\rho\lambda\sigma\tau}
-\frac{15}{4}\mathbf{R}^{2}\mathbf{R}^{2}.\label{18eJan2010}
\end{eqnarray}
Comparing the above two equations by referring to
(\ref{18cJan2010}) , we found
\begin{equation}
S_{\alpha\lambda\sigma\tau}K_{\beta}{}^{\lambda\sigma\tau}
\equiv\frac{1}{4}g_{\alpha\beta}S_{\rho\lambda\sigma\tau}K^{\rho\lambda\sigma\tau}.\label{20aJan2010}
\end{equation}
Using (\ref{20aJan2010}) and considering the quadratic of
$B_{\alpha\beta\mu\nu}$ in (\ref{29Oct2008}), we obtained
\begin{eqnarray}
B_{\alpha\lambda\sigma\tau}B_{\beta}{}^{\lambda\sigma\tau}
&\equiv&-\frac{1}{4}S_{\alpha\lambda\sigma\tau}S_{\beta}{}^{\lambda\sigma\tau}
+K_{\alpha\lambda\sigma\tau}K_{\beta}{}^{\lambda\sigma\tau}
-\frac{15}{32}g_{\alpha\beta}\mathbf{R}^{2}\mathbf{R}^{2},\label{1aDec2008}\\
B_{\rho\lambda\sigma\tau}B^{\rho\lambda\sigma\tau}
&\equiv&-\frac{1}{4}S_{\rho\lambda\sigma\tau}S^{\rho\lambda\sigma\tau}
+K_{\rho\lambda\sigma\tau}K^{\rho\lambda\sigma\tau}
-\frac{15}{8}\mathbf{R}^{2}\mathbf{R}^{2}.\label{1bDec2008}
\end{eqnarray}
Comparing these two equations by using the identities
({\ref{11aJan2010}}) and (\ref{18cJan2010}), we found
\begin{equation}
K_{\alpha\lambda\sigma\tau}K_{\beta}{}^{\lambda\sigma\tau}
\equiv\frac{1}{4}g_{\alpha\beta}K_{\rho\lambda\sigma\tau}K^{\rho\lambda\sigma\tau}.\label{23aJan2010}
\end{equation}
Expand this identity explicitly:
\begin{eqnarray}
K_{\alpha\lambda\sigma\tau}K_{\beta}{}^{\lambda\sigma\tau}
&\equiv&2R_{\alpha\xi\lambda\kappa}R_{\sigma}{}^{\xi}{}_{\tau}{}^{\kappa}
(R_{\beta\mu}{}^{\lambda}{}_{\nu}R^{\sigma\mu\tau\nu}+
R_{\beta\mu}{}^{\lambda}{}_{\nu}R^{\tau\mu\sigma\nu})+\frac{9}{16}g_{\alpha\beta}\mathbf{R}^{2}\mathbf{R}^{2},\\
K_{\rho\lambda\sigma\tau}K^{\rho\lambda\sigma\tau}
&\equiv&2R_{\rho\xi\lambda\kappa}R_{\sigma}{}^{\xi}{}_{\tau}{}^{\kappa}
(R^{\rho}{}_{\mu}{}^{\lambda}{}_{\nu}R^{\sigma\mu\tau\nu}+
R^{\rho}{}_{\mu}{}^{\lambda}{}_{\nu}R^{\tau\mu\sigma\nu})+\frac{9}{4}g_{\alpha\beta}\mathbf{R}^{2}\mathbf{R}^{2}.
\end{eqnarray}
Using the result in (\ref{23aJan2010}), the above two expressions
can be simplified as
\begin{equation}
R_{\alpha\xi\lambda\kappa}R_{\sigma}{}^{\xi}{}_{\tau}{}^{\kappa}
(R_{\beta\mu}{}^{\lambda}{}_{\nu}R^{\sigma\mu\tau\nu}+R_{\beta\mu}{}^{\lambda}{}_{\nu}R^{\tau\mu\sigma\nu})
\equiv\frac{1}{4}g_{\alpha\beta}R_{\rho\xi\lambda\kappa}R_{\sigma}{}^{\xi}{}_{\tau}{}^{\kappa}
(R^{\rho}{}_{\mu}{}^{\lambda}{}_{\nu}R^{\sigma\mu\tau\nu}+R^{\rho}{}_{\mu}{}^{\lambda}{}_{\nu}R^{\tau\mu\sigma\nu}).
\label{23bJan2010}
\end{equation}
Rewriting (\ref{23bJan2010}) in an abbreviated notation,
\begin{equation}
\tilde{K}_{\alpha\lambda\sigma\tau}\tilde{K}_{\beta}{}^{\lambda\sigma\tau}
\equiv\frac{1}{4}g_{\alpha\beta}\tilde{K}_{\rho\lambda\sigma\tau}\tilde{K}^{\rho\lambda\sigma\tau}.
\end{equation}
Likewise, we found the quadratic $V_{\alpha\beta\mu\nu}$ identity
to be
\begin{equation}
V_{\alpha\lambda\sigma\tau}V_{\beta}{}^{\lambda\sigma\tau}
=\frac{1}{4}g_{\alpha\beta}V_{\rho\lambda\sigma\tau}V^{\rho\lambda\sigma\tau}.
\end{equation}
Moreover, we found some more identities in a similar way:
\begin{eqnarray}
B_{\alpha\lambda\sigma\tau}K_{\beta}{}^{\lambda\sigma\tau}
&\equiv&\frac{1}{4}g_{\alpha\beta}B_{\rho\lambda\sigma\tau}K^{\rho\lambda\sigma\tau},\quad~~{}
B_{\alpha\lambda\sigma\tau}V_{\beta}{}^{\lambda\sigma\tau}
\equiv\frac{1}{4}g_{\alpha\beta}B_{\rho\lambda\sigma\tau}V^{\rho\lambda\sigma\tau},\\
S_{\alpha\lambda\sigma\tau}V_{\beta}{}^{\lambda\sigma\tau}
&\equiv&\frac{1}{4}g_{\alpha\beta}S_{\rho\lambda\sigma\tau}V^{\rho\lambda\sigma\tau},\quad\quad{}
K_{\alpha\lambda\sigma\tau}V_{\beta}{}^{\lambda\sigma\tau}
\equiv\frac{1}{4}g_{\alpha\beta}K_{\rho\lambda\sigma\tau}V^{\rho\lambda\sigma\tau}.
\end{eqnarray}
We list all the results in a single formula as follows:
\begin{equation}
X_{\alpha\lambda\sigma\tau}Y_{\beta}{}^{\lambda\sigma\tau}
\equiv\frac{1}{4}g_{\alpha\beta}X_{\rho\lambda\sigma\tau}Y^{\rho\lambda\sigma\tau},
\end{equation}
for all $X,Y\in\{B,S,K,V\}$. More fundamentally, we found
\begin{equation}
\tilde{X}_{\alpha\lambda\sigma\tau}\tilde{Y}_{\beta}{}^{\lambda\sigma\tau}
\equiv\frac{1}{4}g_{\alpha\beta}
\tilde{X}_{\rho\lambda\sigma\tau}\tilde{Y}^{\rho\lambda\sigma\tau},\label{20bJan2010}
\end{equation}
for all
$\tilde{X},\tilde{Y}\in\{\tilde{B},\tilde{S},\tilde{K},\tilde{T}\}$.
Bear in mind the symmetry of
$\tilde{X}_{\alpha\beta\mu\nu}=\tilde{X}_{(\alpha\beta)(\mu\nu)}=\tilde{X}_{(\mu\nu)(\alpha\beta)}$.
There comes a question whether the quadratic one-quarter identity
which is shown in (\ref{20bJan2010}) requires some kind of
symmetry property?  The answer is no and we will discuss this in
section 4.

\section{Quadratic one-quarter metric identity}
Expand (\ref{11aJan2010}) in an explicit form
\begin{eqnarray}
B_{\alpha\lambda\sigma\tau}B_{\beta}{}^{\lambda\sigma\tau}
&\equiv&2R_{\alpha\xi\lambda\kappa}R_{\sigma}{}^{\xi}{}_{\tau}{}^{\kappa}
(R_{\beta\mu}{}^{\lambda}{}_{\nu}R^{\sigma\mu\tau\nu}
+R_{\beta\mu}{}^{\tau}{}_{\nu}R^{\sigma\mu\lambda\nu})
-\frac{1}{16}g_{\alpha\beta}\mathbf{R}^{2}\mathbf{R}^{2},\label{11bJan2010}\\
B_{\rho\lambda\sigma\tau}B^{\rho\lambda\sigma\tau}
&\equiv&2R_{\rho\xi\lambda\kappa}R_{\sigma}{}^{\xi}{}_{\tau}{}^{\kappa}
(R^{\rho}{}_{\mu}{}^{\lambda}{}_{\nu}R^{\sigma\mu\tau\nu}
+R^{\rho}{}_{\mu}{}^{\tau}{}_{\nu}R^{\sigma\mu\lambda\nu})
-\frac{1}{4}\mathbf{R}^{2}\mathbf{R}^{2}.\label{11cJan2010}
\end{eqnarray}
Using the result in (\ref{11aJan2010}), we found the following
relationship from (\ref{11bJan2010}) and (\ref{11cJan2010}):
\begin{eqnarray}
&&R_{\alpha\xi\lambda\kappa}R_{\sigma}{}^{\xi}{}_{\tau}{}^{\kappa}
(R_{\beta\mu}{}^{\lambda}{}_{\nu}R^{\sigma\mu\tau\nu}
+R_{\beta\mu}{}^{\tau}{}_{\nu}R^{\sigma\mu\lambda\nu})\nonumber\\
&\equiv&\frac{1}{4}g_{\alpha\beta}
R_{\rho\xi\lambda\kappa}R_{\sigma}{}^{\xi}{}_{\tau}{}^{\kappa}
(R^{\rho}{}_{\mu}{}^{\lambda}{}_{\nu}R^{\sigma\mu\tau\nu}
+R^{\rho}{}_{\mu}{}^{\tau}{}_{\nu}R^{\sigma\mu\lambda\nu}).
\end{eqnarray}
There comes a natural question: whether the following are true
independently,
\begin{eqnarray}
R_{\alpha\xi\lambda\kappa}R_{\sigma}{}^{\xi}{}_{\tau}{}^{\kappa}
R_{\beta\mu}{}^{\lambda}{}_{\nu}R^{\sigma\mu\tau\nu}
&\equiv&\frac{1}{4}g_{\alpha\beta}
R_{\rho\xi\lambda\kappa}R_{\sigma}{}^{\xi}{}_{\tau}{}^{\kappa}
R^{\rho}{}_{\mu}{}^{\lambda}{}_{\nu}R^{\sigma\mu\tau\nu},\label{14mJan2010}\\
R_{\alpha\xi\lambda\kappa}R_{\sigma}{}^{\xi}{}_{\tau}{}^{\kappa}
R_{\beta\mu}{}^{\tau}{}_{\nu}R^{\sigma\mu\lambda\nu}
&\equiv&\frac{1}{4}g_{\alpha\beta}
R_{\rho\xi\lambda\kappa}R_{\sigma}{}^{\xi}{}_{\tau}{}^{\kappa}
R^{\rho}{}_{\mu}{}^{\tau}{}_{\nu}R^{\sigma\mu\lambda\nu}.\label{21cJan2010}
\end{eqnarray}
Indeed they are, and this was verified by Edgar and Wingbrant
\cite{Edgar}. Here we suggest another and perhaps an easier way to
obtain this result and some other similar results.

The representations of the quadratic Bel-Robinson tensor are not
unique, as is shown in (\ref{11bJan2010}) and (\ref{11cJan2010}).
Because $B_{\alpha\beta\mu\nu}$ is completely symmetric, we found
some more different expressions, including
\begin{eqnarray}
B_{\alpha\lambda\sigma\tau}B_{\beta}{}^{\lambda\sigma\tau}
&\equiv&2R_{\alpha\xi\lambda\kappa}R_{\sigma}{}^{\xi}{}_{\tau}{}^{\kappa}
(R_{\beta\mu}{}^{\sigma}{}_{\nu}R^{\lambda\mu\tau\nu}
+R_{\beta\mu}{}^{\sigma}{}_{\nu}R^{\tau\mu\lambda\nu})\label{15aJan2010}\\
&\equiv&2R_{\alpha\xi\lambda\kappa}R_{\sigma}{}^{\xi}{}_{\tau}{}^{\kappa}
(R_{\beta\mu}{}^{\sigma}{}_{\nu}R^{\lambda\mu\tau\nu}
+R_{\beta\mu}{}^{\tau}{}_{\nu}R^{\lambda\mu\sigma\nu})\label{15bJan2010}\\
&\equiv&2R_{\alpha\xi\lambda\kappa}R_{\sigma}{}^{\xi}{}_{\tau}{}^{\kappa}
(R_{\beta\mu}{}^{\tau}{}_{\nu}R^{\lambda\mu\sigma\nu}
+R_{\beta\mu}{}^{\lambda}{}_{\nu}R^{\tau\mu\sigma\nu})
-\frac{1}{32}g_{\alpha\beta}\mathbf{R}^{2}\mathbf{R}^{2}\label{15cJan2010}\\
&\equiv&2R_{\alpha\xi\lambda\kappa}R_{\sigma}{}^{\xi}{}_{\tau}{}^{\kappa}
(R_{\beta\mu}{}^{\lambda}{}_{\nu}R^{\sigma\mu\tau\nu}
+R_{\beta\mu}{}^{\tau}{}_{\nu}R^{\sigma\mu\lambda\nu})
-\frac{1}{16}g_{\alpha\beta}\mathbf{R}^{2}\mathbf{R}^{2}.
\end{eqnarray}
The corresponding contracted expressions are
\begin{eqnarray}
B_{\rho\lambda\sigma\tau}B^{\rho\lambda\sigma\tau}
&\equiv&2R_{\rho\xi\lambda\kappa}R_{\sigma}{}^{\xi}{}_{\tau}{}^{\kappa}
(R^{\rho}{}_{\mu}{}^{\sigma}{}_{\nu}R^{\lambda\mu\tau\nu}
+R^{\rho}{}_{\mu}{}^{\sigma}{}_{\nu}R^{\tau\mu\lambda\nu})
\label{14aJan2010}\\
&\equiv&2R_{\rho\xi\lambda\kappa}R_{\sigma}{}^{\xi}{}_{\tau}{}^{\kappa}
(R^{\rho}{}_{\mu}{}^{\sigma}{}_{\nu}R^{\lambda\mu\tau\nu}
+R^{\rho}{}_{\mu}{}^{\tau}{}_{\nu}R^{\lambda\mu\sigma\nu})\label{14bJan2010}\\
&\equiv&2R_{\rho\xi\lambda\kappa}R_{\sigma}{}^{\xi}{}_{\tau}{}^{\kappa}
(R^{\rho}{}_{\mu}{}^{\tau}{}_{\nu}R^{\lambda\mu\sigma\nu}
+R^{\rho}{}_{\mu}{}^{\lambda}{}_{\nu}R^{\tau\mu\sigma\nu})
-\frac{1}{8}\mathbf{R}^{2}\mathbf{R}^{2}\label{14cJan2010}\\
&\equiv&2R_{\rho\xi\lambda\kappa}R_{\sigma}{}^{\xi}{}_{\tau}{}^{\kappa}
(R^{\rho}{}_{\mu}{}^{\lambda}{}_{\nu}R^{\sigma\mu\tau\nu}
+R^{\rho}{}_{\mu}{}^{\tau}{}_{\nu}R^{\sigma\mu\lambda\nu})
-\frac{1}{4}\mathbf{R}^{2}\mathbf{R}^{2}.\label{14gJan2010}
\end{eqnarray}
Examining the four pairs of equations (\ref{15aJan2010}) and
(\ref{15bJan2010}), (\ref{14aJan2010}) and (\ref{14bJan2010}),
(\ref{15bJan2010}) and (\ref{15cJan2010}), (\ref{14bJan2010}) and
(\ref{14cJan2010}), we found
\begin{eqnarray}
R_{\alpha\xi\lambda\kappa}R_{\sigma}{}^{\xi}{}_{\tau}{}^{\kappa}
R_{\beta\mu}{}^{\sigma}{}_{\nu}R^{\tau\mu\lambda\nu}
&\equiv&R_{\alpha\xi\lambda\kappa}R_{\sigma}{}^{\xi}{}_{\tau}{}^{\kappa}
R_{\beta\mu}{}^{\tau}{}_{\nu}R^{\lambda\mu\sigma\nu},\label{14eJan2010}\\
R_{\rho\xi\lambda\kappa}R_{\sigma}{}^{\xi}{}_{\tau}{}^{\kappa}
R^{\rho}{}_{\mu}{}^{\sigma}{}_{\nu}R^{\tau\mu\lambda\nu}
&\equiv&R_{\rho\xi\lambda\kappa}R_{\sigma}{}^{\xi}{}_{\tau}{}^{\kappa}
R^{\rho}{}_{\mu}{}^{\tau}{}_{\nu}R^{\lambda\mu\sigma\nu},\label{17aFeb2010}\\
R_{\alpha\xi\lambda\kappa}R_{\sigma}{}^{\xi}{}_{\tau}{}^{\kappa}
R_{\beta\mu}{}^{\sigma}{}_{\nu}R^{\lambda\mu\tau\nu}
&\equiv&R_{\alpha\xi\lambda\kappa}R_{\sigma}{}^{\xi}{}_{\tau}{}^{\kappa}
R_{\beta\mu}{}^{\lambda}{}_{\nu}R^{\tau\mu\sigma\nu}
-\frac{1}{32}g_{\alpha\beta}\mathbf{R}^{2}\mathbf{R}^{2},\\
R_{\rho\xi\lambda\kappa}R_{\sigma}{}^{\xi}{}_{\tau}{}^{\kappa}
R^{\rho}{}_{\mu}{}^{\sigma}{}_{\nu}R^{\lambda\mu\tau\nu}
&\equiv&R_{\rho\xi\lambda\kappa}R_{\sigma}{}^{\xi}{}_{\tau}{}^{\kappa}
R^{\rho}{}_{\mu}{}^{\lambda}{}_{\nu}R^{\tau\mu\sigma\nu}-\frac{1}{8}\mathbf{R}^{2}\mathbf{R}^{2}.
\label{14fJan2010}
\end{eqnarray}
Note that (\ref{14eJan2010}) and (\ref{17aFeb2010}) are trivial
equalities, because it can be obtained from renaming the dummy
indices.

Here we list out the two explicit quadratic expressions of
$S_{\alpha\beta\mu\nu}$:
\begin{eqnarray}
S_{\alpha\lambda\sigma\tau}S_{\beta}{}^{\lambda\sigma\tau}
&\equiv&2R_{\alpha\lambda\xi\kappa}R_{\sigma\tau}{}^{\xi\kappa}(R_{\beta}{}^{\lambda}{}_{\mu\nu}R^{\sigma\tau\mu\nu}
+R_{\beta}{}^{\tau}{}_{\mu\nu}R^{\sigma\lambda\mu\nu})
+\frac{1}{2}g_{\alpha\beta}\mathbf{R}^{2}\mathbf{R}^{2},\label{21aJan2010}\\
S_{\rho\lambda\sigma\tau}S^{\rho\lambda\sigma\tau}
&\equiv&2R_{\rho\lambda\xi\kappa}R_{\sigma\tau}{}^{\xi\kappa}(R^{\rho\lambda}{}_{\mu\nu}R^{\sigma\tau\mu\nu}
+R^{\rho\tau}{}_{\mu\nu}R^{\sigma\lambda\mu\nu})+2\mathbf{R}^{2}\mathbf{R}^{2}.\label{21bJan2010}
\end{eqnarray}
Using (\ref{18cJan2010}), the relationship between
(\ref{21aJan2010}) and (\ref{21bJan2010}) becomes
\begin{eqnarray}
&&R_{\alpha\lambda\xi\kappa}R_{\sigma\tau}{}^{\xi\kappa}(R_{\beta}{}^{\lambda}{}_{\mu\nu}R^{\sigma\tau\mu\nu}
+R_{\beta}{}^{\tau}{}_{\mu\nu}R^{\sigma\lambda\mu\nu})\nonumber\\
&\equiv&\frac{1}{4}g_{\alpha\beta}
R_{\rho\lambda\xi\kappa}R_{\sigma\tau}{}^{\xi\kappa}(R^{\rho\lambda}{}_{\mu\nu}R^{\sigma\tau\mu\nu}
+R^{\rho\tau}{}_{\mu\nu}R^{\sigma\lambda\mu\nu}).
\end{eqnarray}
As before, like the situation of Edgar and Wingbrant \cite{Edgar},
one may wonder whether the following are true independently for
any frames
\begin{eqnarray}
R_{\alpha\lambda\xi\kappa}R_{\sigma\tau}{}^{\xi\kappa}R_{\beta}{}^{\lambda}{}_{\mu\nu}R^{\sigma\tau\mu\nu}
&\equiv&\frac{1}{4}g_{\alpha\beta}
R_{\rho\lambda\xi\kappa}R_{\sigma\tau}{}^{\xi\kappa}R^{\rho\lambda}{}_{\mu\nu}R^{\sigma\tau\mu\nu},
\label{23aApril2010}\\
R_{\alpha\lambda\xi\kappa}R_{\sigma\tau}{}^{\xi\kappa}R_{\beta}{}^{\tau}{}_{\mu\nu}R^{\sigma\lambda\mu\nu}
&\equiv&\frac{1}{4}g_{\alpha\beta}
R_{\rho\lambda\xi\kappa}R_{\sigma\tau}{}^{\xi\kappa}R^{\rho\tau}{}_{\mu\nu}R^{\sigma\lambda\mu\nu}.
\end{eqnarray}
Once again, we found they are indeed true, having verified these
relations in orthonormal frames.  Moreover, using symmetry
properties, we also obtained
\begin{eqnarray}
R_{\alpha\lambda\xi\kappa}R_{\sigma\tau}{}^{\xi\kappa}R_{\beta}{}^{\tau}{}_{\mu\nu}R^{\sigma\lambda\mu\nu}
&\equiv&R_{\alpha\lambda\xi\kappa}R_{\sigma\tau}{}^{\xi\kappa}R_{\beta}{}^{\sigma}{}_{\mu\nu}R^{\lambda\tau\mu\nu},\\
R_{\rho\lambda\xi\kappa}R_{\sigma\tau}{}^{\xi\kappa}R^{\rho\tau}{}_{\mu\nu}R^{\sigma\lambda\mu\nu}
&\equiv&R_{\rho\lambda\xi\kappa}R_{\sigma\tau}{}^{\xi\kappa}R^{\rho\sigma}{}_{\mu\nu}R^{\lambda\tau\mu\nu}.
\end{eqnarray}
Based on the first Bianchi identity, we found the following
relations:
\begin{eqnarray}
R_{\alpha\xi\lambda\kappa}R_{\sigma}{}^{\xi}{}_{\tau}{}^{\kappa}
(R_{\beta\mu}{}^{\lambda}{}_{\nu}R^{\sigma\mu\tau\nu}
-R_{\beta\mu}{}^{\lambda}{}_{\nu}R^{\tau\mu\sigma\nu})
&\equiv&\frac{1}{8}R_{\alpha\lambda\xi\kappa}R_{\sigma\tau}{}^{\xi\kappa}
R_{\beta}{}^{\lambda}{}_{\mu\nu}R^{\sigma\tau\mu\nu},
\label{14nJan2010}\\
R_{\rho\xi\lambda\kappa}R_{\sigma}{}^{\xi}{}_{\tau}{}^{\kappa}
(R^{\rho}{}_{\mu}{}^{\lambda}{}_{\nu}R^{\sigma\mu\tau\nu}
-R^{\rho}{}_{\mu}{}^{\lambda}{}_{\nu}R^{\tau\mu\sigma\nu})
&\equiv&\frac{1}{8}R_{\rho\lambda\xi\kappa}
R_{\sigma\tau}{}^{\xi\kappa}R^{\rho\lambda}{}_{\mu\nu}R^{\sigma\tau\mu\nu}.
\label{14oJan2010}
\end{eqnarray}
Referring to (\ref{23bJan2010}), (\ref{23aApril2010}),
(\ref{14nJan2010}) and (\ref{14oJan2010}), we found
\begin{eqnarray}
R_{\alpha\xi\lambda\kappa}R_{\sigma}{}^{\xi}{}_{\tau}{}^{\kappa}R_{\beta\mu}{}^{\lambda}{}_{\nu}R^{\sigma\mu\tau\nu}
&\equiv&\frac{1}{4}g_{\alpha\beta}
R_{\rho\xi\lambda\kappa}R_{\sigma}{}^{\xi}{}_{\tau}{}^{\kappa}R^{\rho}{}_{\mu}{}^{\lambda}{}_{\nu}R^{\sigma\mu\tau\nu},
\\
R_{\alpha\xi\lambda\kappa}R_{\sigma}{}^{\xi}{}_{\tau}{}^{\kappa}R_{\beta\mu}{}^{\lambda}{}_{\nu}R^{\tau\mu\sigma\nu}
&\equiv&\frac{1}{4}g_{\alpha\beta}
R_{\rho\xi\lambda\kappa}R_{\sigma}{}^{\xi}{}_{\tau}{}^{\kappa}R^{\rho}{}_{\mu}{}^{\lambda}{}_{\nu}R^{\tau\mu\sigma\nu}.
\end{eqnarray}
Consider the difference of the two terms on the right hand side of
(\ref{14aJan2010})
\begin{eqnarray}
&&R_{\rho\xi\lambda\kappa}R_{\sigma}{}^{\xi}{}_{\tau}{}^{\kappa}
R^{\rho}{}_{\mu}{}^{\sigma}{}_{\nu}R^{\lambda\mu\tau\nu}
-R_{\rho\xi\lambda\kappa}R_{\sigma}{}^{\xi}{}_{\tau}{}^{\kappa}
R^{\rho}{}_{\mu}{}^{\sigma}{}_{\nu}R^{\tau\mu\lambda\nu}\nonumber\\
&\equiv&\frac{1}{16}(2R_{\rho\lambda\xi\kappa}R_{\sigma\tau}{}^{\xi\kappa}
R^{\rho\lambda}{}_{\mu\nu}R^{\sigma\tau\mu\nu}
+4R_{\rho\lambda\xi\kappa}R_{\sigma\tau}{}^{\xi\kappa}
R^{\rho\tau}{}_{\mu\nu}R^{\sigma\lambda\mu\nu}-\mathbf{R}^{2}\mathbf{R}^{2})\equiv{}0,\label{20dJan2010}
\end{eqnarray}
where we have found and made use of the following identity, which
was verified by using orthonormal frames:
\begin{equation}
2R_{\rho\lambda\xi\kappa}R_{\sigma\tau}{}^{\xi\kappa}
R^{\rho\lambda}{}_{\mu\nu}R^{\sigma\tau\mu\nu}
+4R_{\rho\lambda\xi\kappa}R_{\sigma\tau}{}^{\xi\kappa}
R^{\rho\tau}{}_{\mu\nu}R^{\sigma\lambda\mu\nu}\equiv\mathbf{R}^{2}\mathbf{R}^{2}.\label{17bFeb2010}
\end{equation}
Using the result in (\ref{20dJan2010}), we noted that
(\ref{14aJan2010}) can be rewritten as
\begin{equation}
\frac{1}{4}B_{\rho\lambda\sigma\tau}B^{\rho\lambda\sigma\tau}
\equiv{}R_{\rho\xi\lambda\kappa}R_{\sigma}{}^{\xi}{}_{\tau}{}^{\kappa}
R^{\rho}{}_{\mu}{}^{\sigma}{}_{\nu}R^{\lambda\mu\tau\nu}
\equiv{}R_{\rho\xi\lambda\kappa}R_{\sigma}{}^{\xi}{}_{\tau}{}^{\kappa}
R^{\rho}{}_{\mu}{}^{\sigma}{}_{\nu}R^{\tau\mu\lambda\nu}.\label{31cMar2010}
\end{equation}
Using (\ref{31cMar2010}), refer to (\ref{14cJan2010}), we
discovered
\begin{equation}
\frac{1}{4}B_{\rho\lambda\sigma\tau}B^{\rho\lambda\sigma\tau}
\equiv{}R_{\rho\xi\lambda\kappa}R_{\sigma}{}^{\xi}{}_{\tau}{}^{\kappa}
R^{\rho}{}_{\mu}{}^{\lambda}{}_{\nu}R^{\tau\mu\sigma\nu}
-\frac{1}{16}\mathbf{R}^{2}\mathbf{R}^{2}.
\end{equation}
In order to take care of the first two terms on the right hand
side of (\ref{14gJan2010}) in a similar way, we need to use
another identity which comes from Deser \cite{DeserarXiv}
\begin{equation}
2B_{\rho\lambda\sigma\tau}B^{\rho\lambda\sigma\tau}
\equiv\mathbf{R}^{2}\mathbf{R}^{2}
-2R_{\rho\lambda\xi\kappa}R_{\sigma\tau}{}^{\xi\kappa}R^{\rho\lambda}{}_{\mu\nu}R^{\sigma\tau\mu\nu}.
\label{14hJan2010}
\end{equation}
We have checked the above identity using orthonormal frames,
however, our result presented here differs from \cite{DeserarXiv}
by the coefficient of `2' at the last term in (\ref{14hJan2010}).
According to (\ref{14gJan2010}) and substituting this amazing
identity in (\ref{14hJan2010}), we obtained
\begin{eqnarray}
R_{\rho\xi\lambda\kappa}R_{\sigma}{}^{\xi}{}_{\tau}{}^{\kappa}
R^{\rho}{}_{\mu}{}^{\lambda}{}_{\nu}R^{\sigma\mu\tau\nu}
&\equiv&\frac{1}{8}B_{\rho\lambda\sigma\tau}B^{\rho\lambda\sigma\tau}
+\frac{1}{8}\mathbf{R}^{2}\mathbf{R}^{2},\\
R_{\rho\xi\lambda\kappa}R_{\sigma}{}^{\xi}{}_{\tau}{}^{\kappa}
R^{\rho}{}_{\mu}{}^{\tau}{}_{\nu}R^{\sigma\mu\lambda\nu}
&\equiv&\frac{3}{8}B_{\rho\lambda\sigma\tau}B^{\rho\lambda\sigma\tau}.
\end{eqnarray}

There are only four fourth rank quadratic-in-Weyl-curvature
tensors having the specified symmetry; they were mentioned
previously: (\ref{20aJan2009}) to (\ref{20bJan2009}). From this it
follows that there exits only two independent
quadratic-in-Weyl-curavture scalar expressions.  Explicitly
\begin{equation}
R_{\rho\xi\lambda\kappa}R_{\sigma}{}^{\xi}{}_{\tau}{}^{\kappa}
R^{\rho}{}_{\mu}{}^{\sigma}{}_{\nu}R^{\lambda\mu\tau\nu},\quad{}
\mathbf{R}^{2}\mathbf{R^{2}},\label{19hMar2010}
\end{equation}
(for this result see also \cite{DeserarXiv}). The others are just
the linear combinations of these two.  For example
\begin{eqnarray}
R_{\rho\xi\lambda\kappa}R_{\sigma}{}^{\xi}{}_{\tau}{}^{\kappa}
R^{\rho\tau}{}_{\mu\nu}R^{\sigma\lambda\mu\nu}
&=&2R_{\rho\xi\lambda\kappa}R_{\sigma}{}^{\xi}{}_{\tau}{}^{\kappa}
(R^{\rho}{}_{\mu}{}^{\tau}{}_{\nu}R^{\sigma\mu\lambda\nu}
-R^{\rho}{}_{\mu}{}^{\tau}{}_{\nu}R^{\lambda\mu\sigma\nu})\nonumber\\
&=&R_{\rho\xi\lambda\kappa}R_{\sigma}{}^{\xi}{}_{\tau}{}^{\kappa}
R^{\rho}{}_{\mu}{}^{\sigma}{}_{\nu}R^{\lambda\mu\tau\nu}.
\end{eqnarray}
After going through some messy algebra, the two basis components
for the square quadratic curvature tensors which were mentioned in
(\ref{19hMar2010}) denote a simple fact.  There exists an identity
(i.e., not a condition) such that
\begin{equation}
X_{\alpha\lambda\sigma\tau}Y_{\beta}{}^{\lambda\sigma\tau}
\equiv\frac{1}{4}g_{\alpha\beta}X_{\rho\lambda\sigma\tau}Y^{\rho\lambda\sigma\tau},\label{31bMar2010}
\end{equation}
where $X_{\alpha\lambda\sigma\tau}$ and
$Y_{\beta\lambda\sigma\tau}$ are any tensors quadratic in the
Riemann curvature, non-vanishing in vacuum.

Looking back at the identity in (\ref{14hJan2010}), there comes a
completeness question.  As the Bel-Robinson tensor satisfies the
dominant energy condition, which means the sign of
$B_{\alpha\beta\mu\nu}u^{\alpha}v^{\beta}w^{\mu}z^{\nu}$ is
non-negative. Does there exist a definite sign of the quadratic
Bel-Robinson tensor?  Checking the sign of
$B^{2}_{\alpha\beta\mu\nu}$, we used the five distinct Petrov
types \cite{Petrov} as a verification technique in orthonormal
frames, we found
\begin{equation}
B_{\alpha\beta\mu\nu}B^{\alpha\beta\mu\nu}\geq{}0.
\end{equation}
This result indicates that it is true for all frames because
$B_{\alpha\beta\mu\nu}$ is a tensor.   Alternatively, using the
$(3+1)$ decomposition and the identity in (\ref{17bFeb2010}), we
recovered the same result
\begin{eqnarray}
B_{\rho\lambda\sigma\tau}B^{\rho\lambda\sigma\tau}
&=&\frac{1}{4}\mathbf{R}^{2}\mathbf{R}^{2}
+\frac{2}{3}(R_{0a\xi\kappa}R_{bc}{}^{\xi\kappa}+R_{0b\xi\kappa}R_{ca}{}^{\xi\kappa}+R_{0c\xi\kappa}R_{ab}{}^{\xi\kappa})\nonumber\\
&&\times(R_{0}{}^{a}{}_{\mu\nu}R^{bc\mu\nu}+R_{0}{}^{b}{}_{\mu\nu}R^{ca\mu\nu}
+R_{0}{}^{c}{}_{\mu\nu}R^{ab\mu\nu})\geq0.
\end{eqnarray}

For the completeness, as $B_{\alpha\beta\mu\nu}$ and
$V_{\alpha\beta\mu\nu}$ share the same value of the
energy-momentum density in vacuum, one may wonder what is the sign
of the quadratic of $V_{\alpha\beta\mu\nu}$?  We found that,
unfortunately, the sign is not certain. Here is the simple
derivation
\begin{eqnarray}
V_{\rho\lambda\sigma\tau}V^{\rho\lambda\sigma\tau}
&=&(B_{\rho\lambda\sigma\tau}+W_{\rho\lambda\sigma\tau})(B^{\rho\lambda\sigma\tau}+W^{\rho\lambda\sigma\tau})\nonumber\\
&=&B_{\rho\lambda\sigma\tau}B^{\rho\lambda\sigma\tau}
+2B_{\rho\lambda\sigma\tau}W^{\rho\lambda\sigma\tau}+W_{\rho\lambda\sigma\tau}W^{\rho\lambda\sigma\tau}\nonumber\\
&=&\frac{1}{8}(9\mathbf{R}^{2}\mathbf{R}^{2}-10B_{\rho\lambda\sigma\tau}B^{\rho\lambda\sigma\tau}).\label{27aApril2010}
\end{eqnarray}
where
\begin{eqnarray}
B_{\rho\lambda\sigma\tau}W^{\rho\lambda\sigma\tau}=0,\quad{}
W_{\rho\lambda\sigma\tau}W^{\rho\lambda\sigma\tau}
=\frac{9}{8}(\mathbf{R}^{2}\mathbf{R}^{2}-2B_{\rho\lambda\sigma\tau}B^{\rho\lambda\sigma\tau}).
\end{eqnarray}
In Petrov type II \cite{Petrov}, (\ref{27aApril2010}) becomes
\begin{equation}
V_{\rho\lambda\sigma\tau}V^{\rho\lambda\sigma\tau}=9[13(E^{2}_{11}-H^{2}_{11})^{2}-20E^{2}_{11}H^{2}_{11}].
\end{equation}
Note that if either $E_{11}$ is much larger than $H_{11}$ or
conversely, $V_{\rho\lambda\sigma\tau}V^{\rho\sigma\tau}$ is
positive.  However, if $E_{11}$ is very close to $H_{11}$, then
the sign of $V_{\rho\lambda\sigma\tau}V^{\rho\lambda\sigma\tau}$
will become negative.  Hence the sign of
$V_{\rho\lambda\sigma\tau}V^{\rho\lambda\sigma\tau}$ is not
certain.

\section{Algebraic Rainich conditions}
The original algebraic Rainich conditions use the Ricci tensor.
Making an analogy from the second rank to a fourth rank traceless
tensor, we write
\begin{equation}
X_{\alpha\lambda\sigma\tau}X_{\beta}{}^{\lambda\sigma\tau}
=\frac{1}{4}g_{\alpha\beta}X_{\rho\lambda\sigma\tau}X^{\rho\lambda\sigma\tau},~~{}
0=X^{\alpha}{}_{\alpha\mu\nu}=X^{\alpha}{}_{\mu\alpha\nu}=...,~~{}
X_{\alpha\beta\mu\nu}u^{\alpha}u^{\beta}u^{\mu}u^{\nu}\geq0,\label{21dJan2010}
\end{equation}
where $u$ is timelike unit normal vector (the latter relation is
equivalent to $X_{0000}\geq0$).  Note that we assumed the
coefficient of $X_{\alpha\beta\mu\nu}$ is positive. The basic idea
of the algebraic Rainich conditions do not require the dominant
energy condition, but just the weak energy condition
\cite{Stephani}.  For the fourth rank tensor, as far as the
quasilocal in the small sphere limit is concerned, we found that
only $B_{\alpha\beta\mu\nu}$ and $V_{\alpha\beta\mu\nu}$ satisfy
these algebraic Rainich conditions. Moreover, we modify the
algebraic Rainich conditions as follows:
\begin{equation}
0=X^{\alpha}{}_{\alpha\mu\nu}=X^{\alpha}{}_{\mu\alpha\nu}=...,\quad{}
X_{\alpha\beta\mu\nu}u^{\alpha}u^{\beta}u^{\mu}u^{\nu}\geq0,
\end{equation}
where the first requirement in (\ref{21dJan2010}) is ignored
because it is an identity but not a condition, which is explained
in section 4.  Here we found that the completely traceless
property of $X_{\alpha\beta\mu\nu}$ gives the basis of
$B_{\alpha\beta\mu\nu}$ and $V_{\alpha\beta\mu\nu}$, and these two
tensors imply positivity (more precisely inside the forward light
cone, briefly causal). Interestingly, this is also true
conversely. Therefore the fourth rank algebraic Rainich conditions
can be further simplified. In short
\begin{equation}
0=X^{\alpha}{}_{\alpha\mu\nu}=X^{\alpha}{}_{\mu\alpha\nu}=...\quad\Leftrightarrow\quad
X_{\alpha\beta\mu\nu}u^{\beta}u^{\mu}u^{\nu}=(E_{ab}E^{ab}+H_{ab}H^{ab},2\epsilon_{cab}E^{ad}H^{b}{}_{d}).
\end{equation}
This indicates that, as far as the quasilocal small sphere limit,
the algebraic Rainich conditions only require one condition.  In
other words, either the completely traceless or positivity (i.e.,
causal) is sufficient.  The following is the simple proof.

Case (i).  Completely traceless property implies positivity (more
precisely causal). Recall the four basic tensors from
(\ref{20aJan2009}) to (\ref{20bJan2009}), because of the
symmetries of $\tilde{B}_{\alpha\beta\mu\nu}$,
$\tilde{S}_{\alpha\beta\mu\nu}$, $\tilde{K}_{\alpha\beta\mu\nu}$
and $\tilde{T}_{\alpha\beta\mu\nu}$, consider the two following
totally traceless statements:
\begin{eqnarray}
0=a_{1}\tilde{B}^{\alpha}{}_{\alpha\mu\nu}
+a_{2}\tilde{S}^{\alpha}{}_{\alpha\mu\nu}
+a_{3}\tilde{K}^{\alpha}{}_{\alpha\mu\nu}
+a_{4}\tilde{T}^{\alpha}{}_{\alpha\mu\nu}
=\frac{1}{2}(a_{1}+a_{2}-a_{4})g_{\mu\nu}\mathbf{R}^{2},
\label{17cFeb2010}\\
0=a_{1}\tilde{B}^{\alpha}{}_{\mu\alpha\nu}
+a_{2}\tilde{S}^{\alpha}{}_{\mu\alpha\nu}
+a_{3}\tilde{K}^{\alpha}{}_{\mu\alpha\nu}
+a_{4}\tilde{T}^{\alpha}{}_{\mu\alpha\nu}
=\frac{1}{8}(a_{1}-2a_{2}+3a_{3}-a_{4})g_{\mu\nu}\mathbf{R}^{2}.
\label{17dFeb2010}
\end{eqnarray}
Then we have two constraints,
\begin{eqnarray}
&&0=a_{1}+a_{2}-a_{4},\\
&&0=a_{1}-2a_{2}+3a_{3}-a_{4}.
\end{eqnarray}
The solution for the above two equations can be represented as
\begin{equation}
a_{4}=a_{1}+a_{2},\quad{}a_{2}=a_{3}.
\end{equation}
Then the general linear combination of the four basic fundamental
tensors (i.e., confined in the quasilocal small region) can be
reduced as
\begin{eqnarray}
&&a_{1}\tilde{B}_{\alpha\beta\mu\nu}+a_{2}\tilde{S}_{\alpha\beta\mu\nu}
+a_{3}\tilde{K}_{\alpha\beta\mu\nu}+a_{4}\tilde{T}_{\alpha\beta\mu\nu}\nonumber\\
&=&a_{1}(\tilde{B}_{\alpha\beta\mu\nu}+\tilde{T}_{\alpha\beta\mu\nu})
+a_{2}(\tilde{S}_{\alpha\beta\mu\nu}+\tilde{K}_{\alpha\beta\mu\nu}+\tilde{T}_{\alpha\beta\mu\nu})\nonumber\\
&=&a_{1}B_{\alpha\beta\mu\nu}+a_{2}V_{\alpha\beta\mu\nu}.
\end{eqnarray}
This result indicates that there are only two tensors
$B_{\alpha\beta\mu\nu}$ and $V_{\alpha\beta\mu\nu}$ which satisfy
the completely trace free property and form a linear basis.
Obviously, they also fulfill the positivity (i.e., causal)
\begin{equation}
B_{\mu{}000}=V_{\mu{}000}=(E_{ab}E^{ab}+H_{ab}H^{ab},~2\epsilon_{cab}E^{ad}H^{b}{}^{d}),
\end{equation}
where the energy and momentum density represent the causal
relationship:
\begin{equation}
E_{ab}E^{ab}+H_{ab}H^{ab}\geq|2\epsilon_{cab}E^{ad}H^{b}{}_{d}|\geq{}0.\label{14aApril2010}
\end{equation}

Case (ii).  Positivity (more precisely causal) implies completely
traceless.  First of all, why do we keep emphasizing the
positivity and causal? The reason is that positivity alone cannot
imply completely trace free.  In particular, suppose
\begin{eqnarray}
X_{\alpha\beta\mu\nu}=R_{\alpha\lambda\beta\sigma}R_{\mu}{}^{\lambda}{}_{\nu}{}^{\sigma}.
\end{eqnarray}
Although $X_{\alpha\beta\mu\nu}$ preserves the positive condition,
it does not satisfy the completely traceless property. Explicitly
\begin{eqnarray}
X^{\alpha}{}_{\mu\alpha\nu}=\frac{1}{4}g_{\mu\nu}\mathbf{R}^{2}\neq0,\quad{}
X_{0000}=E_{ab}E^{ab}\geq0.
\end{eqnarray}
Returning back to causal, consider the energy-momentum integral in
a quasilocal small sphere with constant time evolution of the
hypersurface. Note that the fourth rank tensor
$X_{\alpha\beta\mu\nu}$ needs to be symmetric at the last two
indices because of the small sphere limit
\begin{eqnarray}
N^{\mu}P_{\mu}&=&\int{}N^{\mu}X^{\rho}{}_{\mu\xi\kappa}x^{\xi}x^{\kappa}\eta_{\rho}
=\int{}N^{\mu}X^{0}{}_{\mu\xi\kappa}x^{\xi}x^{\kappa}\eta_{0}
=\int{}N^{\mu}X^{0}{}_{\mu{}ij}x^{i}x^{j}dV\nonumber\\
&=&\int{}N^{\mu}X^{0}{}_{\mu{}l}{}^{l}\frac{r^{2}}{3}dV
=\int{}N^{\mu}X^{0}{}_{\mu{}l}{}^{l}\frac{r^{2}}{3}4\pi{}r^{2}dr
=N^{\mu}X^{0}{}_{\mu{}l}{}^{l}\frac{4\pi{}r^{5}}{15}\nonumber\\
&=&N^{\mu}\left(X^{0}{}_{\mu\alpha}{}^{\alpha}-X^{0}{}_{\mu{}0}{}^{0}\right)\frac{4\pi{}r^{5}}{15}
=N^{\mu}X^{0}{}_{\mu{}00}\frac{4\pi{}r^{5}}{15},
\end{eqnarray}
where we made the assumption that $X_{0\mu\alpha}{}^{\alpha}$
vanishes and fulfills causal (i.e., Lorentz-covariant, see section
4.2.2 of \cite{Szabados}). Consider now the requirement for the
energy-momentum being future pointing and non-spacelike (i.e.,
causal) in the small sphere limit :
\begin{eqnarray}
&&a_{1}\tilde{B}_{\mu0l}{}^{l}+a_{2}\tilde{S}_{\mu0l}{}^{l}
+a_{3}\tilde{K}_{\mu0l}{}^{l}+a_{4}\tilde{T}_{\mu0l}{}^{l}\nonumber\\
&=&a_{1}(-2E_{ab}E^{ab}+4H_{ab}H^{ab},~2\epsilon_{cab}E^{ad}H^{b}{}_{d})
+a_{2}(-4E_{ab}E^{ab}+4H_{ab}H^{ab},~0)\nonumber\\
&&+a_{3}(2E_{ab}E^{ab},~2\epsilon_{cab}E^{ad}H^{b}{}_{d})
+a_{4}(3E_{ab}E^{ab}-3H_{ab}H^{ab},~0)\nonumber\\
&=&(-2a_{1}-4a_{2}+2a_{3}+3a_{4})E_{ab}E^{ab}+(4a_{1}+4a_{2}-3a_{4})H_{ab}H^{ab}\nonumber\\
&&+(2a_{1}+2a_{3})\epsilon_{cab}E^{ad}H^{b}{}_{d}.\label{12aApril2010}
\end{eqnarray}
Causal (i.e., Lorentz-covariant, see \cite{Szabados}) requires the
magnitude of $E_{ab}E^{ab}$ and $H_{ab}H^{ab}$ to be the same and
the energy is greater than or equal to the momentum as shown in
(\ref{14aApril2010}). Simply, we need to calculate two equations
from (\ref{12aApril2010}), but it turns out that one constraint is
enough, requiring the coefficients of the electric and magnetic
square parts to be the same
\begin{equation}
-2a_{1}-4a_{2}+2a_{3}+3a_{4}=4a_{1}+4a_{2}-3a_{4},
\end{equation}
Then $a_{4}$ can be written in terms of $a_{1}$, $a_{2}$ and
$a_{3}$
\begin{equation}
a_{4}=a_{1}+\frac{4a_{2}}{3}-\frac{a_{3}}{3}.\label{12bApril2010}
\end{equation}
Substituting (\ref{12bApril2010}) into (\ref{12aApril2010}), we
found
\begin{eqnarray}
&&a_{1}\tilde{B}_{\mu0l}{}^{l}+a_{2}\tilde{S}_{\mu0l}{}^{l}
+a_{3}\tilde{K}_{\mu0l}{}^{l}+a_{4}\tilde{T}_{\mu0l}{}^{l}\nonumber\\
&=&a_{1}(\tilde{B}_{\mu0l}{}^{l}+\tilde{T}_{\mu0l}{}^{l})
+a_{2}\left(\tilde{S}_{\mu0l}{}^{l}+\frac{4}{3}\tilde{T}_{\mu0l}{}^{l}\right)
+a_{3}\left(\tilde{K}_{\mu0l}{}^{l}-\frac{1}{3}\tilde{T}_{\mu0l}{}^{l}\right)\nonumber\\
&=&a_{1}B_{\mu0l}{}^{l}+a_{3}\left(\tilde{K}_{\mu000}-\frac{1}{3}\tilde{T}_{\mu0l}{}^{l}\right)\nonumber\\
&=&a_{1}B_{\mu000}+a_{3}(\tilde{K}_{\mu000}
+\tilde{S}_{\mu000}+\tilde{T}_{\mu000})\nonumber\\
&=&a_{1}B_{\mu000}+a_{3}V_{\mu000}\nonumber\\
&=&(a_{1}+a_{3})B_{\mu{}000},
\end{eqnarray}
where we require $a_{1}+a_{3}\geq0$ and made the following
substitutions
\begin{eqnarray}
\tilde{S}_{\mu0l}{}^{l}&=&-\frac{4}{3}\tilde{T}_{\mu0l}{}^{l},\label{13aApril2010}\\
B_{\mu0\alpha}{}^{\alpha}&=&0~=~\tilde{K}_{\mu0\alpha}{}^{\alpha},\\
B_{\alpha\beta\mu\nu}&=&\tilde{B}_{\alpha\beta\mu\nu}+\tilde{T}_{\alpha\beta\mu\nu},\\
V_{\alpha\beta\mu\nu}&=&\tilde{S}_{\alpha\beta\mu\nu}+\tilde{K}_{\alpha\beta\mu\nu}+\tilde{T}_{\alpha\beta\mu\nu},
\end{eqnarray}
and we also used
\begin{equation}
-\frac{1}{3}\tilde{T}_{\mu{}0l}{}^{l}=\tilde{S}_{\mu{}000}+\tilde{T}_{\mu{}000},
\end{equation}
which can easily be verified.

Hence, the completely traceless and causal properties form
necessary and sufficient conditions.  This means we can further
simplify the algebraic Rainich conditions for a fourth rank
tensor; as far as the quasilocal small sphere limit is concerned,
we only need the completely trace free condition or positivity
(i.e., causal).  This is an interesting result which is valid in
the quasilocal small sphere region.

Moreover, we have found another interesting result for the general
case (i.e., not confined to the quasilocal small sphere limit)
which will be discussed in the next section.

\section{Algebraic Rainich conditions for general fourth rank tensor}
For the fourth rank tensors, so far we only confined ourselves in
the quasilocal small sphere limit, $B_{\alpha\beta\mu\nu}$ and
$V_{\alpha\beta\mu\nu}$. What about the general situation for this
fourth rank tensor?  Generally speaking, we found remarkably that
the algebraic Rainich conditions only require the completely trace
free property.  This means the totally traceless condition
automatically fulfills the positivity; however, the converse does
not apply.

In principle, assuming vacuum (i.e., the Ricci tensor vanishes),
using
$R_{\alpha\beta\mu\nu}=R_{[\alpha\beta][\mu\nu]}=R_{\mu\nu\alpha\beta}$
we can get the eighteen combinations for the fourth rank quadratic
Riemann curvature tensors:\\

\begin{tabular}{llllll}
\hline
$R_{\alpha\lambda\mu\sigma}R_{\beta}{}^{\lambda}{}_{\nu}{}^{\sigma}$
&$R_{\alpha\mu\lambda\sigma}R_{\beta}{}^{\lambda}{}_{\nu}{}^{\sigma}$
&$R_{\alpha\lambda\mu\sigma}R_{\beta\nu}{}^{\lambda\sigma}$
&$R_{\alpha\mu\lambda\sigma}R_{\beta\nu}{}^{\lambda\sigma}$
&$R_{\alpha\lambda\nu\sigma}R_{\beta}{}^{\lambda}{}_{\mu}{}^{\sigma}$\\
$R_{\alpha\nu\lambda\sigma}R_{\beta}{}^{\lambda}{}_{\mu}{}^{\sigma}$
&$R_{\alpha\lambda\nu\sigma}R_{\beta\mu}{}^{\lambda\sigma}$
&$R_{\alpha\nu\lambda\sigma}R_{\beta\mu}{}^{\lambda\sigma}$
&$R_{\alpha\lambda\mu\sigma}R_{\nu}{}^{\lambda}{}_{\beta}{}^{\sigma}$
&$R_{\alpha\lambda\nu\sigma}R_{\mu}{}^{\lambda}{}_{\beta}{}^{\sigma}$\\
$R_{\alpha\lambda\beta\sigma}R_{\mu}{}^{\lambda}{}_{\nu}{}^{\sigma}$
&$R_{\alpha\beta\lambda\sigma}R_{\mu}{}^{\lambda}{}_{\nu}{}^{\sigma}$
&$R_{\alpha\lambda\beta\sigma}R_{\mu\nu}{}^{\lambda\sigma}$
&$R_{\alpha\beta\lambda\sigma}R_{\mu\nu}{}^{\lambda\sigma}$
&$R_{\alpha\lambda\beta\sigma}R_{\nu}{}^{\lambda}{}_{\mu}{}^{\sigma}$\\
$g_{\alpha\beta}g_{\mu\nu}R^{2}$ &$g_{\alpha\mu}g_{\beta\nu}R^{2}$
&$g_{\alpha\nu}g_{\beta\mu}R^{2}$\\
\hline
\end{tabular}
\begin{center}
Table 1: Eighteen quadratic Riemann curvature tensors
\end{center}
From the first Bianich identity $R_{\alpha[\beta\mu\nu]}=0$ and
the identity (\ref{7Nov2008}), these can be reduced to the
following eight algebraically linearly independent expressions:
\begin{eqnarray}
\begin{array}{cccc}
R_{\alpha\lambda\mu\sigma}R_{\beta}{}^{\lambda}{}_{\nu}{}^{\sigma},
&R_{\alpha\lambda\nu\sigma}R_{\beta}{}^{\lambda}{}_{\mu}{}^{\sigma},
&R_{\alpha\mu\lambda\sigma}R_{\beta\nu}{}^{\lambda\sigma},
&R_{\alpha\nu\lambda\sigma}R_{\beta\mu}{}^{\lambda\sigma},\\
R_{\alpha\lambda\beta\sigma}R_{\mu}{}^{\lambda}{}_{\nu}{}^{\sigma},
&R_{\alpha\lambda\beta\sigma}R_{\nu}{}^{\lambda}{}_{\mu}{}^{\sigma},
&g_{\alpha\mu}g_{\beta\nu}R^{2},
&g_{\alpha\nu}g_{\beta\mu}R^{2}.\\
\end{array}
\end{eqnarray}
Here we consider two cases.\\
Case (i). Completely trace free, without any symmetry requirement.
We define
\begin{eqnarray}
X_{\alpha\beta\mu\nu}&:=&b_{1}R_{\alpha\lambda\mu\sigma}R_{\beta}{}^{\lambda}{}_{\nu}{}^{\sigma}
+b_{2}R_{\alpha\lambda\nu\sigma}R_{\beta}{}^{\lambda}{}_{\mu}{}^{\sigma}
+b_{3}R_{\alpha\mu\lambda\sigma}R_{\beta\nu}{}^{\lambda\sigma}
+b_{4}R_{\alpha\nu\lambda\sigma}R_{\beta\mu}{}^{\lambda\sigma}\nonumber\\
&&+b_{5}R_{\alpha\lambda\beta\sigma}R_{\mu}{}^{\lambda}{}_{\nu}{}^{\sigma}
+b_{6}R_{\alpha\lambda\beta\sigma}R_{\nu}{}^{\lambda}{}_{\mu}{}^{\sigma}
+b_{7}g_{\alpha\mu}g_{\beta\nu}R^{2}
+b_{8}g_{\alpha\nu}g_{\beta\mu}R^{2},\label{19dMar2010}
\end{eqnarray}
where $b_{1}$ to $b_{8}$ are constants.  Basically, in order to
set the totally traceless for $X_{\alpha\beta\mu\nu}$, there
should be six possible combinations. However, we found there are
only three independent constraints. Explicitly
\begin{eqnarray}
&&0=X^{\alpha}{}_{\alpha\mu\nu}=X_{\mu\nu}{}^{\alpha}{}_{\alpha}
=\left(\frac{b_{1}}{4}+\frac{b_{2}}{4}+\frac{b_{3}}{4}+\frac{b_{4}}{4}
+b_{7}+b_{8}\right)g_{\mu\nu}\mathbf{R}^{2},\label{19aMar2010}\\
&&0=X^{\alpha}{}_{\mu\alpha\nu}=X_{\mu}{}^{\alpha}{}_{\nu\alpha}
=\left(\frac{b_{2}}{8}-\frac{b_{4}}{4}+\frac{b_{5}}{4}+\frac{b_{6}}{8}
+4b_{7}+b_{8}\right)g_{\mu\nu}\mathbf{R}^{2},\label{19bMar2010}\\
&&0=X^{\alpha}{}_{\mu\nu\alpha}=X_{\mu}{}^{\alpha}{}_{\alpha\nu}
=\left(\frac{b_{1}}{8}-\frac{b_{3}}{4}+\frac{b_{5}}{8}+\frac{b_{6}}{4}
+b_{7}+4b_{8}\right)g_{\mu\nu}\mathbf{R}^{2}.\label{19cMar2010}
\end{eqnarray}
Using the constraints from (\ref{19aMar2010}) to
(\ref{19cMar2010}), we eliminate $b_{6}$, $b_{7}$, $b_{8}$ and
rewrite (\ref{19dMar2010}) as
\begin{eqnarray}
X_{\alpha\beta\mu\nu}&=&~~b_{1}\left(R_{\alpha\lambda\mu\sigma}R_{\beta}{}^{\lambda}{}_{\nu}{}^{\sigma}
+3R_{\alpha\lambda\beta\sigma}B_{\nu}{}^{\lambda}{}_{\mu}{}^{\sigma}
-\frac{1}{24}g_{\alpha\mu}g_{\beta\nu}\mathbf{R}^{2}
-\frac{5}{24}g_{\alpha\nu}g_{\beta\mu}\mathbf{R}^{2}\right)\nonumber\\
&&+b_{2}\left(R_{\alpha\lambda\nu\sigma}R_{\beta}{}^{\lambda}{}_{\mu}{}^{\sigma}
+3R_{\alpha\lambda\beta\sigma}R_{\nu}{}^{\lambda}{}_{\mu}{}^{\sigma}
-\frac{1}{12}g_{\alpha\mu}g_{\beta\nu}\mathbf{R}^{2}
-\frac{1}{6}g_{\alpha\nu}g_{\beta\mu}\mathbf{R}^{2}
\right)\nonumber\\
&&+b_{3}\left(R_{\alpha\mu\lambda\sigma}R_{\beta\nu}{}^{\lambda\sigma}
+4R_{\alpha\lambda\beta\sigma}R_{\nu}{}^{\lambda}{}_{\mu}{}^{\sigma}
-\frac{1}{12}g_{\alpha\mu}g_{\beta\nu}\mathbf{R}^{2}
-\frac{1}{6}g_{\alpha\nu}g_{\beta\mu}\mathbf{R}^{2}\right)\nonumber\\
&&+b_{4}\left(R_{\alpha\nu\lambda\sigma}R_{\beta\mu}{}^{\lambda\sigma}
+4R_{\alpha\lambda\beta\sigma}R_{\nu}{}^{\lambda}{}_{\mu}{}^{\sigma}
-\frac{1}{4}g_{\alpha\nu}g_{\beta\mu}\mathbf{R}^{2}
\right)\nonumber\\
&&+b_{5}\left(R_{\alpha\lambda\beta\sigma}R_{\mu}{}^{\lambda}{}_{\nu}{}^{\sigma}
-R_{\alpha\lambda\beta\sigma}R_{\nu}{}^{\lambda}{}_{\mu}{}^{\sigma}
-\frac{1}{24}g_{\alpha\mu}g_{\beta\nu}\mathbf{R}^{2}+\frac{1}{24}g_{\alpha\nu}g_{\beta\mu}\mathbf{R}^{2}
\right).\quad
\end{eqnarray}
Note that in general this is not simply a linear combination of
$B_{\alpha\beta\mu\nu}$ and $V_{\alpha\beta\mu\nu}$.  We found
that the totally traceless property implies positivity,
\begin{equation}
0=X^{\alpha}{}_{\alpha\mu\nu}=X^{\alpha}{}_{\mu\alpha\nu}=...\quad{}\Rightarrow\quad{}X_{0000}\geq0.
\end{equation}
as long as $b_{1}$ to $b_{5}$ are all non-negative.  But the
converse is not true in general. In particular
\begin{equation}
X_{\alpha\beta\mu\nu}:=R_{\alpha\lambda\mu\sigma}R_{\beta}{}^{\lambda}{}_{\nu}{}^{\sigma},
\end{equation}
satisfies the positivity requirement
\begin{equation}
X_{0000}=E_{ab}E^{ab}\geq0.
\end{equation}
But does not satisfy the totally traceless property, since
\begin{equation}
X^{\alpha}{}_{\alpha\mu\nu}=R^{\alpha}{}_{\lambda\mu\sigma}R_{\alpha}{}^{\lambda}{}_{\nu}{}^{\sigma}
=\frac{1}{4}g_{\mu\nu}\mathbf{R}^{2}\neq0.
\end{equation}\\
Case (ii). Completely trace free with symmetry requirement.  We
impose one symmetry condition and then allow the totally trace
free condition afterward.  Set
$X'_{\alpha\beta\mu\nu}=X'_{\alpha\beta(\mu\nu)}$ and it will have
the following implications simultaneously
\begin{equation}
X'_{\alpha\beta\mu\nu}=X'_{\alpha\beta(\mu\nu)}\quad\Rightarrow\quad{}
X'_{\alpha\beta\mu\nu}=X'_{(\alpha\beta)\mu\nu}
\quad {\mbox{and}} \quad{}
X'_{\alpha\beta\mu\nu}=X'_{\mu\nu\alpha\beta} .\label{21bMar2010}
\end{equation}
For instance, let
\begin{eqnarray}
X'_{\alpha\beta\mu\nu}
&=&c_{1}(R_{\alpha\lambda\mu\sigma}R_{\beta}{}^{\lambda}{}_{\nu}{}^{\sigma}
+R_{\alpha\lambda\nu\sigma}R_{\beta}{}^{\lambda}{}_{\mu}{}^{\sigma})
+c_{2}(R_{\alpha\mu\lambda\sigma}R_{\beta\nu}{}^{\lambda\sigma}
+R_{\alpha\nu\lambda\sigma}R_{\beta\mu}{}^{\lambda\sigma})\nonumber\\
&&+c_{3}(R_{\alpha\lambda\beta\sigma}R_{\mu}{}^{\lambda}{}_{\nu}{}^{\sigma}
+R_{\alpha\lambda\beta\sigma}R_{\nu}{}^{\lambda}{}_{\mu}{}^{\sigma})
+c_{4}(g_{\alpha\mu}g_{\beta\nu}\mathbf{R}^{2}+g_{\alpha\nu}g_{\beta\mu}\mathbf{R}^{2})\nonumber\\
&=&(c_{1}-8c_{4})\tilde{B}_{\alpha\beta\mu\nu}+(c_{2}-4c_{4})\tilde{S}_{\alpha\beta\mu\nu}
+(c_{3}+8c_{4})\tilde{K}_{\alpha\beta\mu\nu}-16c_{4}\tilde{T}_{\alpha\beta\mu\nu}\nonumber\\
&=&c'_{1}\tilde{B}_{\alpha\beta\mu\nu}+c'_{2}\tilde{S}_{\alpha\beta\mu\nu}
+c'_{3}\tilde{K}_{\alpha\beta\mu\nu}+c'_{4}\tilde{T}_{\alpha\beta\mu\nu},\label{19eMar2010}
\end{eqnarray}
where $c_{1}$ to $c_{4}$ or $c'_{1}$ to $c'_{4}$ are constants,
and we have made use of the property (\ref{7Nov2008}).  In order
to fulfill the completely trace free requirements, there are only
two different constraints we need to consider
\begin{eqnarray}
&&0=X'^{\alpha}{}_{\alpha\mu\nu}=c'_{1}+c'_{2}-c'_{4},\label{19fMar2010}\\
&&0=X'^{\alpha}{}_{\mu\alpha\nu}=c'_{1}-2c'_{2}+3c'_{3}-c'_{4}.\label{19gMar2010}
\end{eqnarray}
The solution for the above two equations are
\begin{equation}
c'_{4}=c'_{1}+c'_{2}, \quad{} c'_{2}=c'_{3}. \label{21aMar2010}
\end{equation}
Using (\ref{21aMar2010}), rewrite (\ref{19eMar2010}) as
\begin{eqnarray}
X'_{\alpha\beta\mu\nu}
&=&c'_{1}(\tilde{B}_{\alpha\beta\mu\nu}+\tilde{T}_{\alpha\beta\mu\nu})
+c'_{2}(\tilde{S}_{\alpha\beta\mu\nu}+\tilde{K}_{\alpha\beta\mu\nu}+\tilde{T}_{\alpha\beta\mu\nu})\nonumber\\
&=&c'_{1}B_{\alpha\beta\mu\nu}+c'_{2}V_{\alpha\beta\mu\nu}.
\end{eqnarray}
Hence, starting from the general completely trace free property,
we have recovered the unique basis $B_{\alpha\beta\mu\nu}$ and
$V_{\alpha\beta\mu\nu}$ in the quasilocal small sphere limit.

This means that if we impose some certain symmetry as indicated in
(\ref{21bMar2010}), we obtained the same result as mentioned in
section 5 but without the quasilocal small sphere limit
restriction.  Explicitly, in general, the purely mathematical
property (i.e., completely trace free) guarantees the physical
requirements \cite{SoarXiv2009} (i.e., energy-momentum
conservation and causal).

Likewise, for the completeness, we get the same result if one sets
$X''_{\alpha\beta\mu\nu}=X''_{\alpha(\beta\mu)\nu}$.

\section{Conclusion}
The Bel-Robinson tensor satisfies the one-quarter quadratic
identity
$B_{\alpha\lambda\sigma\tau}B_{\beta}{}^{\lambda\sigma\tau}
=\frac{1}{4}g_{\alpha\beta}$  $B_{\rho\lambda\sigma\tau}
B^{\rho\lambda\sigma\tau}$.  We found that the tensors
$S_{\alpha\beta\mu\nu}$, $K_{\alpha\beta\mu\nu}$ and
$V_{\alpha\beta\mu\nu}$ also satisfy the same interesting
one-quarter quadratic identity as $B_{\alpha\beta\mu\nu}$ does.
Explicitly
$X_{\alpha\lambda\sigma\tau}Y_{\beta}{}^{\lambda\sigma\tau}
\equiv\frac{1}{4}g_{\alpha\beta}X_{\rho\lambda\sigma\tau}Y^{\rho\lambda\sigma\tau}$,
for all $X,Y\in\{B,S,K,V\}$.  More fundamentally, for any
quadratic Riemann curvature tensors
$\tilde{X}_{\alpha\beta\mu\nu}$ and
$\tilde{Y}_{\alpha\beta\mu\nu}$, we have the same result
$\tilde{X}_{\alpha\lambda\sigma\tau}\tilde{Y}_{\beta}{}^{\lambda\sigma\tau}
=\frac{1}{4}g_{\alpha\beta}\tilde{X}_{\rho\lambda\sigma\tau}\tilde{Y}^{\rho\lambda\sigma\tau}$.
This indicates that this is an identity and no longer a condition.
Therefore the algebraic Rainich conditions left two conditions,
not the original three.

Moreover, under the quasilocal small sphere limit restriction, we
found that there are only two fourth rank tensors
$B_{\alpha\beta\mu\nu}$ and $V_{\alpha\beta\mu\nu}$ forming a
basis for good expressions. Both of them have the completely trace
free and causal properties, these two form necessary and
sufficient conditions. Surprisingly, either completely traceless
or causal can fulfill the algebraic Rainich conditions.

Furthermore, relaxing the quasilocal small sphere limit
restriction and considering the general fourth rank tensor, we
found two remarkable results. One is without any symmetry
requirement: the algebraic conditions only require totally trace
free.  The other is imposing some certain symmetry: we recovered
the same result as in the quasilocal small sphere limit (i.e.,
$B_{\alpha\beta\mu\nu}$ and $V_{\alpha\beta\mu\nu}$).

\section*{Acknowledgment}
This work was supported by NSC 97-2811-M-032-007 and NSC
98-2811-M-008-078.

\end{document}